\documentclass[11pt]{article}

\def\beq{\begin{eqnarray}}    
\def\eeq{\end{eqnarray}}       
\def\nn{\nonumber}                
\usepackage{pgfcore}
\usepackage{amsmath}
\usepackage{graphicx}
\usepackage{amssymb}
\usepackage{amsfonts}
\usepackage{latexsym}

\renewcommand{\Re}{\,\mbox{Re}\,}
\renewcommand{\Im}{\,\mbox{Im}\,}
\def\beq{\begin{eqnarray}}
\def\eeq{\end{eqnarray}}
\def\ln{\,\mbox{ln}\,}
\def\Det{\,\mbox{Det}\,}
\def\det{\,\mbox{det}\,}
\def\tr{\,\mbox{tr}\,}
\def\diag{\,\mbox{diag}\,}
\def\Tr{\,\mbox{Tr}\,}
\def\sTr{\,\mbox{sTr}\,}
\def\Res{\,\mbox{Res}\,}
\renewcommand{\Re}{\,\mbox{Re}\,}
\renewcommand{\Im}{\,\mbox{Im}\,}
\def\lap{\Delta}

\def\al{\alpha}
\def\be{\beta}
\def\ch{\chi}
\def\ga{\gamma}
\def\de{\delta}
\def\vp{\varepsilon}
\def\ep{\epsilon}
\def\ze{\zeta}
\def\io{\iota}
\def\ka{\kappa}
\def\la{\lambda}
\def\na{\nabla}
\def\pa{\partial}
\def\ro{\varrho}

\def\si{\sigma}
\def\om{\omega}
\def\ph{\varphi}
\def\ta{\tau}
\def\th{\theta}
\def\te{\vartheta}
\def\up{\upsilon}
\def\Ga{\Gamma}
\def\De{\Delta}
\def\La{\Lambda}
\def\Si{\Sigma}
\def\Om{\Omega}
\def\Te{\Theta}
\def\Th{\Theta}
\def\Up{\Upsilon}

\def\nn{\nonumber}                

\begin{document}

\def\ln{\,\mbox{ln}\,}              
\def\tr{\,\mbox{tr}\,}               
\def\str{\,\mbox{str}\,}            
\def\Tr{\,\mbox{Tr}\,}            
\def\sTr{\,\mbox{sTr}\,}         
\def\Box{\square}                    
\def\cx{\square}                      
\def\det{\,\mbox{det}\,}          
\def\Det{\,\mbox{Det}\,}         
\def\diag{\,\mbox{diag}}         
\def\Res{\,\mbox{Res}\,}        

\def\FL{ Friedmann-–Lema\^{\i}tre }
\def\Sch{ Schwarzschild }
\def\VVM{ Van Vleck-Morette }
\def \SchDW{ Schwinger-DeWitt }
\def\FL{ Friedmann-–Lema\^{\i}tre }
\def\DR{dimensional regularization}

\renewcommand{\Re}{\,\mbox{Re}\,}       
\renewcommand{\Im}{\,\mbox{Im}\,}       
\def\lap{\Delta}                        

\def\al{\alpha}
\def\be{\beta}
\def\ch{\chi}
\def\ga{\gamma}
\def\de{\delta}
\def\ep{\epsilon}
\def\vp{\varepsilon}
\def\ze{\zeta}
\def\io{\iota}
\def\ka{\kappa}
\def\la{\lambda}
\def\na{\nabla}
\def\pa{\partial}
\def\ro{\varrho}
\def\si{\sigma}
\def\om{\omega}
\def\ph{\varphi}
\def\ta{\tau}
\def\th{\theta}
\def\te{\vartheta}
\def\up{\upsilon}
\def\Ga{\Gamma}
\def\De{\Delta}
\def\La{\Lambda}
\def\Si{\Sigma}
\def\Om{\Omega}
\def\Te{\Theta}
\def\Th{\Theta}
\def\Up{\Upsilon}

\def\SSB{spontaneous symmetry breaking }
\def\QG{ quantum gravity }
\def\B1{ Bianchi-I }
\def\FL{ Friedmann - Lema\^{\i}tre }
\def\GW{ gravitational wave }
\def\HL{ Hubble - Lema\^{\i}tre }
\def\CC{ cosmological constant }
\def\Sch{ Schwarzschild }
\def\EMT{ energy-momentum tensor }


\begin{center}

{\Large One-loop effective action: nonlocal form factors
and renormalization group}
\vskip 6mm

\textbf{Poliane de Morais Teixeira} 
,\quad
\textbf{Ilya L. Shapiro} 
,\quad
\textbf{Tiago G. Ribeiro}  
\vskip 4mm

{\sl
Departamento de F\'{\i}sica, ICE, Universidade Federal de Juiz de Fora
\\
Campus Universit\'{a}rio - Juiz de Fora, 36036-330, MG, Brazil
}

\vskip 2mm
\vskip 2mm

{\it E-mail addresses: \ polianemteixeira@gmail.com
\quad
shapiro@fisica.ufjf.br
\quad
tgribeiro@ice.ufjf.br}

\end{center}
\vskip 12mm

\begin{quotation}

\noindent
{\bf Abstract.}
\
We review and present full detail of the Feynman diagram - based
and heat-kernel method - based calculations of the simplest
nonlocal form factors in the one-loop contributions of a massive
scalar field. The paper has a pedagogical and introductory purposes
and is intended to help the reader in better understanding the
existing literature on the subject. The functional calculations are
based on the solution by Avramidi and Barvinsky \& Vilkovisky
for the heat kernel and are performed in curved spacetime. One
of the important points is that the main structure of non-localities
is the same as in the flat background.
\vskip 2mm

\vskip 1mm


\end{quotation}
\vskip 4mm

\section{Introduction}

The main method of calculating quantum loop corrections in
QFT (quantum field theory) is based on the integration of the
Feynman diagrams in momentum representation. At the same
time, to work in curved space (spacetime), one has to go beyond
this technique, because the global Fourier transformation in curved
space is impossible. There are three different main approaches
to the curved-space calculations. The first is based on expanding
the external metric on the flat background
$g_{\mu\nu}=\eta_{\mu\nu} + h_{\mu\nu}$ and making the
calculations in the flat space, treating $h_{\mu\nu}$ as an external
flat-space field. The covariance and locality of the divergences
make such an approach possible and in many cases useful
\cite{UtDW,ZeldStar71}. The same concerns, in many cases, the
derivation of the finite nonlocal part of the diagrams
\cite{apco,omar2013}.

Another approach to the calculations in curved space is based on
the use of normal coordinates and local momentum representation
\cite{BunPar}. One of the advantages of this method is an explicit
covariance. In some cases, it provides serious technical benefits,
e.g. for deriving the effective potential in the mass-dependent
schemes of renormalization \cite{CorPot,BAES}. At the same time,
since the local momentum representation is essentially based on the
expansion in the vicinity of a single spacetime point, this method is
not well suited for the nonlocal contributions.

Finally, the Schwinger-De Witt technique \cite{DeWitt,bavi85} is
the most efficient way to derive the one-loop divergences in a curved
space background. About 25 years ago there was significant progress
in the development of the heat-kernel methods by Avramidi
\cite{avram}, Barvinsky and Vilkovisky \cite{bavi90}. As a result,
the general expressions for the non-localities in curved space have
been derived, and this opened the way for calculating the one-loop
nonlocal form factors for different fields
\cite{apco,fervi,omar2013,sebastian} and models
(see e.g. \cite{Bexi}).

From the viewpoint of physical applications, the similarities and main
differences between standard Schwinger-De Witt technique and the
new heat-kernel methods are as follows. In both cases, one deals
with the first few terms in the derivative expansion of the covariant
effective action in ``curvatures''. In the case of gravity, due to the
covariance, this expansion has the form of a power series in the
curvature tensor and its contractions (curvatures). Also, for the
operator of the standard form
\beq
\hat{\cal H}\,=\, \hat{1}\cx \,+\, 2\hat{h}^\al \na_\al
\,+\, \hat{\Pi},
\label{Pih}
\eeq
the expressions such as
\beq
\hat{\cal P}\,=\, \hat{\Pi}
\,+\,\frac{\hat{1}}{6}\,R
\,-\,\na_\al \hat{h}^\al
\,-\,\hat{h}_\al \hat{h}^\al
\label{P}
\eeq
and
\beq
\hat{\cal S}_{\al\be}\,=\, \big[ \na_\be ,\,\na_\al \big]\hat{1}
\,+\,\na_\be \hat{h}_\al
\,-\,\na_\al  \hat{h}_\be
\,+\,\hat{h}_\be \hat{h}_\al
\,-\,\hat{h}_\al \hat{h}_\be
\label{S}
\eeq
are also included in the list of curvatures.

Many physical applications are based on the terms which are
quadratic and at most cubic in curvatures. The main difference is
that the standard Schwinger-De Witt technique deals with the high
energy limit (namely, it is related to the limit $s\rightarrow 0$ in
the proper-time representation). The corresponding terms are
UV-divergent and hence local. As a result, they are usually
irrelevant in the IR limit. Of course, for massless fields, there is
a certain duality between UV and IR,
hence one can always restore the most important part of the
IR-relevant nonlocal terms, e.g., by integrating conformal anomaly.
However,  in the case of the massive field such integration can not
be used or it has a very restricted physical sense \cite{Shocom,asta},
because of the IR decoupling in gravity. In general, the decoupling
is important since it enables one to separate the relevant and
irrelevant degrees of freedom at low energies (in the IR) and thus it
represents one of the main ingredients of the effective field theory
approach.

At least one of the first papers on the gravitational decoupling was
\cite{ZeldStar71}, where it was shown that in the $k^2\ll m^2$ and
$\vert R_{...}\vert \ll m^2$ limit the expression for the renormalized
$\langle T_{\mu\nu}\rangle$  of a massive scalar field in curved
space-time becomes local. The corresponding terms have mass
dependence $\sim m^{-2}$ and no $\mu$-dependence since there
is no direct relation to the UV divergences.
The explicit expressions for the non-local form factors enable one
to explore the details of the IR limit for the massive fields and
hence one can observe and explore such a phenomenon as the
low-energy decoupling.

In what follows we present the details of deriving the gravitational
form-factors, which lead to the gravitational analog of the Appelquist
and Carazzone decoupling theorem \cite{AC}.

Indeed, the heat-kernel solution of \cite{bavi90} is known only
for the operators of the form (\ref{Pih}), while in some cases we
need to work with the operators of different form, where the solution
for the heat-kernel is unknown. Thus, it is very important to establish
the relation between the Feynman diagrams - based and heat-kernel
based calculations of the nonlocal form factors. There are some
calculations of this sort in the particular cases\footnote{There was
also earlier calculation of the gravitational form factors with
temperature \cite{GuZeln}, but without analysis of decoupling.}
\cite{apco,sebastian}, but they deal with the specific cases of the
free fields on curved background and are technically complicated.
For this reason, we present a very simple, pedagogical derivation of
the nonlocal form factors in flat space using diagrams and compare
it with the heat-kernel calculation, which just repeats (correcting
some misprints) the one of \cite{apco,fervi}. For the sake of
generality, the diagram calculation is partially performed in
dimensional regularization and in the covariant Euclidean cut-off
regularization, demonstrating the equivalence between the two
regularizations for the logarithmic divergences, something being
certainly well-known in the different contexts (see e.g.
\cite{Salam51,bavi85,Liao,CorPot}). Our purpose is to present
this known feature in the clear and simple form.

The paper is organized as follows. In the next Sec.~\ref{regstypes}
we discuss the calculation of divergences and nonlocal form factors
in dimensional and the covariant cut-off regularizations. In
Sec.~\ref{nonloca} we demonstrate the general derivation of
nonlocal form factors in curved spacetime, in full detail. In
Sec.~\ref{nlscal} the example of a massive scalar field is
elaborated. We do not go into similar detail for the massive
fermions and vector fields, because it could be quite boring and,
also because the reader can easily elaborate these two examples
as exercises, e.g. following the papers \cite{fervi,omar2013,sebastian}.
Finally, in Sec.~\ref{Conc} we draw our conclusions and discuss
some perspectives in this area.

\section{Two types of the UV regularization}
\label{regstypes}

There are many different regularization schemes and in fact, it is
not difficult to invent a new one. The most used examples are cut-off
regularizations (including three-dimensional cut-off in momentum
space, four-dimensional Euclidean and the covariant cut-off in the
proper-time integral), Pauli-Villars (conventional, covariant and
higher-derivative covariant), analytic (different versions),
zeta-regularization (which is not a regularization, properly
speaking), point-splitting and the dimensional regularization, which
has advantages to preserve the gauge symmetry and be the simplest
one, in many cases.

Since the dimensional regularization \cite{leibr} preserves the gauge
symmetry explicitly (unlike cut-off and some others) it can be used
not only at the one-loop order but also for the multi-loop diagrams.
The disadvantages are that one can not see quadratic divergences,
also it is not really ``physical'', such as the cut-off regularization,
for instance. Anyway, dimensional regularization is one of the most
used regularizations, hence let us describe its use in detail.

\subsubsection{Mathematical preliminaries}
\label{sect2}

We shall need a few special mathematical tools, as reviewed  below.
\vskip 2mm

\noindent
\textbf{1. Analytic continuation.} \
Consider two regions $D_1$ and $D_2$  on the complex plane.
Analytic continuation theorem tells us that in some cases one can
extend the analytic function from some set of points to the larger
region, uniquely.

Consider the two functions $F_1(z)$ and $F_2(z)$, defined and
analytic on  $D_1$ and $D_2$, correspondingly. Suppose
$D_1 \cap D_2 = D$. Furthermore, we assume that $F_1(z)=F_2(z)$
on a set which belongs to $D$ and has at least one accumulation
point. Then, $F_1(z)=F_2(z)$ on the whole $D$.

Our strategy will be to define such continuation for the badly
defined integrals (in Euclidean signature), such as e.g.
\beq
I_4 &=& \int \frac{d^4p }{(2\pi)^4}\,\,
\frac{1}{(p^2 + m^2)\big[(p-k)^2 + m^2\big]}.
\label{I}
\eeq
We assume this integral to be defined in Euclidean four-dimensional
space, but our purpose is to make a continuation from dimension four
to a complex dimension 2$\om$, $I_{4} \to I_{2\om}$, such that
$I_{2\om}$ is analytic on the complex plane except same at most
countable number of points. Then, in the vicinity of the point
$\om=2$ we have
\beq
I_{2\om} = \Big(\textrm{divergent} \sim \frac{1}{2-\om} \;
\textrm{pole term} \Big)\,+\,\mbox{finite terms}\,
+\, \mbox{vanishing} \,\, O(2-\om) \,\, \mbox{terms}.
\nn
\eeq
Our first purpose will be to establish the first, divergent term,
with the pole at $\,\om=2$.

\vskip 2mm

\noindent
\textbf{2. Gaussian integral.} \

This integral in the dimension $\,2\om\,$ reads
\beq
\int \frac{d^{2\om}k}{(2\pi)^{2\om}} \,e^{-xk^2 + 2kb}
&=& \frac{1}{(2\pi)^{2\om}} \Big( \frac{\pi}{x} \Big)^{\om}
e^{\frac{b^2}{x}}.
\label{Gau}
\eeq
For a natural $2\om=1,2,3,4,\dots$, this integral can be easily
derived. For complex values of $\om$, Eq.~(\ref{Gau}) should
be seen as a definition, or as analytic continuation. One can see
the standard review \cite{leibr} for detailed explanation of the
procedure of continuation $\,4 \to n \to 2\om$.

A typical example of applying (\ref{Gau}) is related to the
representation
\beq
\frac{1}{k^2+m^2}
&=&
\int_0 ^\infty d\alpha \; e^{-\alpha(k^2+m^2)}.
\label{Feyn}
\eeq
Consider the continuation of the integral (\ref{I}) into
dimension $n=2\om$,
\beq
I_{2\om}
&=& \int
\frac{d^{2\om}k}{(2\pi)^{2\om}(k^2 + m^2)\big[(k-p)^2 + m^2\big]}
\nn
\\
&=& \int \frac{d^{2\om}k}{(2\pi)^{2\om}} \int_0 ^\infty d\alpha_1
\int_0 ^\infty d\alpha_2 \;
e^{-\alpha_1(k^2+m^2) - \alpha_2[(k-p)^2+m^2]}.
\label{I2om}
\eeq
Changing the order of integrations, it is easy to note that the
integral over $\,k\,$ is exactly of the type (\ref{Gau}), hence
we arrive at
\beq
I_{2\om}
&=&
\int_0 ^\infty d\alpha_1 \int_0 ^\infty d\alpha_2 \int
\frac{d^{2\om}k}{(2\pi)^{2\om}} \;
e^{-k^2(\alpha_1 + \alpha_2) + 2\alpha_2kp
- (\alpha_1+\alpha_2)m^2 - \alpha_2p^2}
\nn
\\
&=&
\int_0 ^\infty d\alpha_1 \int_0 ^\infty d\alpha_2 \;
\frac{1}{(2\pi)^{2\om}} \Big( \frac{\pi}{\al_1 + \al_2}\Big)^\om
e^{\frac{\al_2^2 p^2}{\al_1+\al_2} - \alpha_2(p^2+m^2)
- \al_1m^2}.
\eeq
\\
The last representation will prove useful, at some moment.

\vskip 2mm

\noindent
\textbf{3. Some properties of the gamma-function.}

The gamma-function is defined as
\beq
\Ga{(z)} &=& \int_0 ^\infty dt \; t^{z-1} e^{-t} .
\eeq
The main properties which concern us, are
\beq
&&
\Ga{(z+1)} = z\Ga{(z)} \,\,\, \Longrightarrow \,\,\,
\Ga{(n+1)} = n !\,,
\nn
\\
&&
\Ga{\Big( \frac{1}{2} \Big)} = \sqrt{\pi}
\,\,\,\Longrightarrow\,\,\,
\Ga{\Big(n+ \frac{1}{2} \Big)}
= \frac{1\cdot3\cdot5 \cdots}{2^n} \sqrt{\pi}\,,
\nn
\\
&& \Ga{(z)}\,=\, \lim_{n\rightarrow \infty}
\frac{n ! \; n^z}{z(z+1)\cdots(z+n)}.
\label{Ga}
\eeq
From the last representation directly follows that $\Ga{(z)}$
has simple poles at zero and the negative integer points,
$z=0,-1,-2,\dots$, and nowhere else. Another representation, where
this fact can be seen explicitly, is the Weirstrass's partial fraction
expansion
\beq
\Ga{(z)}\,=\, \Ga_n{(z)} &=& \sum_{n=0}^\infty
\frac{(-1)^n}{n ! (n+z)}
+ \int_1^\infty dt \; t^{z-1} e^{-t} .
\label{GaW}
\eeq
It is clear that $\Ga{(z)}$ is analytic everywhere except
$z=0,-1,-2,\dots$.

An explicit representation of $\Ga{(1-\om)}$ can be obtained
from
\beq
\Ga{(2-\om)} =  (1-\om)\Ga{(1-\om)}
\label{Ga2}
\eeq
and Eq.~(\ref{Ga}),
\beq
\Ga{(2-\om)}
\,=\, \lim_{n \rightarrow \infty} J_\om ,
\quad
\mbox{where}
\quad
J_\om
\,=\, \frac{n ! \; n^{2-\om}} {(2-\om)(3-\om)\cdots(n+2-\om)}.
\label{Gamma2}
\eeq
The expression under the limit can be transformed as
\beq
J_\om &=&
\frac{n ! \; e^{(2-\om)\ln n}}{(2-\om)(1+2-\om)(2+2-\om)
\cdots(n+2-\om)}.
\nn
\eeq
Obviously, the divergent part of $J_\om$, in the limit $\om \to 2$, is
\beq
J_\om^{(div)} &=& \frac{ n ! \cdot1}{(2-\om)\cdot1 \cdot2 \cdots n}
= \frac{1}{2-\om}.
\nn
\eeq
The finite part can be evaluated by means of the following
transformations:
\beq
\frac{1}{2-\om} \; e^{(2-\om)\ln n}
&=& \frac{1}{2-\om} \Big[1 + (2-\om)\ln{n}
+ {\cal O}\big((2-\om)^2\big) \Big]
\nn
\\
&=&\frac{1}{2-\om} + \ln{n} + {\cal O}(2-\om) \;,
\nn
\\
\frac{1}{k-\om} &=& \frac{1}{k-2+(2-\om)}
\,=\, \frac{1}{k-2} \cdot \frac{1}{1+\frac{2-\om}{k-2}}
\,=\, \frac{1}{k-2} \Big(1 - \frac{2-\om}{k-2}+ \cdots \Big) .
\nn
\eeq
Therefore,
\beq
J_\om
&=&
\frac{1}{2-\om} + \ln{n} - \Big( 1 + \frac{1}{2}+ \cdots
+ \frac{1}{n} \Big) + {\cal O}(2-\om) .
\eeq
The sum of the finite terms is
\beq
\ga
&=&
\lim_{n \to \infty} \bigg(1 + \frac{1}{2} + \cdots
+ \frac{1}{n} -  \ln{n} \bigg).
\eeq
and its value is
$\gamma=0,57721\dots$ (Euler-Mascheroni, or just Euler's constant).

Regardless of we shall keep it, the finite contribution has no much
importance, because it sums up with an infinite term
$\frac{1}{2-\om}$. The conventional notation is
\beq
\frac{1}{\vp} \;=\; \frac{1}{(4 \pi)^2 (n-4)} ,
\qquad
n-4\;=\; - 2(2-\om)\,.
\nn
\eeq
Finally, from (\ref{Gamma2}) and (\ref{Ga2}) follows
\beq
&&
 \Gamma (2-\om) \,=\,
 \int_0^\infty \frac{e^{-t}\,dt }{t^{1-w}}
\,=\,\frac{1}{2-\om} -\gamma + {\cal O}(2-\om) \,,
\label{gama2}
\\
&&
\Ga{(1-\om)} = \frac{1}{1-\om} \Ga{(2-\om)}
= \frac{1}{-1+(2-\om)} \Ga{(2-\om)}
= -\frac{1}{2-\om} -1 +\gamma + {\cal O}(2-\om),
\mbox{\qquad}
\label{gamma17}
\\
&&
\Ga{(-\om)} = \frac{1}{2(2-\om)}
+ \frac{3}{4} - \frac{\ga}{2} + {\cal O}(2-\om)\,.
\label{Ga3}
\eeq

\vskip 2mm

\noindent
\textbf{4. Volume of the sphere.} \

Finally, let us calculate the volume of the $m$-dimensional sphere
with the radius
\beq
R=(x_1^2 + x_2^2 + \cdots + x_m^2)^{1/2}.
\nn
\eeq
The dimensional arguments tell us that
\beq
V_m &=& C_m R^m,
\label{V}
\eeq
where $C_m$ are the coefficients which we need to calculate.
For this sake, consider the Gaussian integral
\beq
I &=& \int\limits_{-\infty}^\infty  dx_1
\int\limits_{-\infty}^\infty  dx_2
\cdots \int\limits_{-\infty}^\infty dx_m
\; e^{-a(x_1^2 + x_2^2 + \cdots + x_m^2)}
\,\,\bigg[ \int_{-\infty}^\infty dx \; e^{-ax^2} \bigg]^m
= \Big( \frac{\pi}{a} \Big)^{\frac{m}{2}}.
\label{II}
\eeq
On the other hand, $dV_m = mC_mR^{m-1}dR$, hence
\beq
I &=& \int\limits_{-\infty}^\infty  e^{-aR^2}mC_m\,R^{m-1}\, dR.
\nn
\eeq
Making the change of variables $z=aR^2$, we get
\beq
dR = \frac{1}{2a}\Big(\frac{a}{z} \Big)^{\frac{1}{2}} dz,
&& R^{m-1}= \frac{z}{a}^{\frac{m-1}{2}}
\nn
\eeq
and therefore
\beq
I \,=\, mC_m \int_0^\infty e^{-z}
\Big( \frac{z}{a} \Big)^\frac{m-1}{2} \cdot
\frac{1}{2a} \Big(\frac{a}{z} \Big)^\frac{1}{2} dz
\,=\, \frac{mC_m}{2a^\frac{m}{2}} \int_0^\infty e^{-z}
z^{\frac{m}{2} - 1} dz
\,=\,
 \frac{mC_m}{2a^\frac{m}{2}} \; \Ga{\Big(\frac{m}{2} \Big)}.
 \mbox{\quad}
\label{Im}
\eeq
Since (\ref{II}) and (\ref{Im}) is the same thing, we get
\beq
C_m
= \frac{\pi^\frac{m}{2}}{\frac{m}{2} \Ga{\left( \frac{m}{2} \right)}}
= \frac{\pi^\frac{m}{2}}{\Ga{\left( \frac{m}{2} + 1 \right)}}
\qquad \Longrightarrow \qquad
V_m &=& \frac{\pi^\frac{m}{2}}{\Ga{\left( \frac{m}{2} + 1 \right)}}
\; R^m .
\eeq
The last relation is valid for any natural $m$, but we can also
continue it to an arbitrary complex dimension $2\om$.

\subsubsection{The simplest loop integral}
\label{sect3}

Now we are in a position to start regularizing loop integrals.
The general strategy will be to continue
\beq
I_4 \quad \rightarrow \quad I_{2\om}
&=& \int \frac{d^{2\om}x}{(2\pi)^{2\om}} \cdots\,,
\eeq
such that $I_{2\om}$ is defined in all complex plane except same
points, including $\om=2$. Typically,
\beq
I_{2\om} = (\text{pole at } \om=2) + \text{regular terms}.
\nn
\eeq
\vskip 1mm

Consider the scalar theory with the $\la \ph^4$ interaction,
\beq
S
&=&
\int d^4 z\, \Big\{\frac{1}{2}(\partial \ph)^2 - \frac{m^2}{2} \ph^2
- \frac{\la}{4 !} \ph^4 \Big\}.
\eeq
Let us first rewrite it as Euclidean action, by changing the variable
$z^0=-iz^4$. Then
\beq
&&
d^4z = dz^0 d^3z = -idz^4 d^3z = -id^4z_E
\nn
\\
\mbox{and}
\,\,\,
&&
(\partial \ph)^2 = (\partial_0 \ph)^2  -(\partial \ph)^2 =
- (\partial \ph)_E^2.
\eeq
Finally,
\beq
S
&=&
-i \int d^4 z_E \,\Big\{
-\frac{1}{2} (\pa \ph)_E^2
- \frac{m^2}{2}\ph^2 -\frac{\la}{4 !}\ph^4 \Big\}.
\eeq
Consider the diagram
\beq
\frac12\,\,
\includegraphics[keepaspectratio=true,width=2.0cm,height=2cm]
{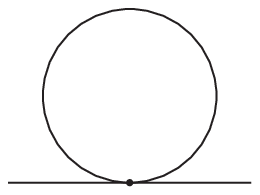}
\,\,=\,\,
\frac12\,I_{go}
\,\,=\,\,
- \frac{\la}{2}
\int \frac{d^4p}{(2\pi)^4} \; \frac{1}{p^2+m^2}\,.
\eeq

First of all, consider the cut-off calculation of this simple diagram,
\beq
&&
\frac12\,I_{go}
\,\,=\,\,
- \,\frac{\la}{2} \cdot \frac{1}{16\pi^2} \cdot \frac{\pi^2}{2}
\int_0^\Om \frac{4p^3dp}{p^2+m^2}
\,=\, -\frac{\la}{32\pi^2} \int_0^\Om \frac{p^2 dp^2}{p^2+m^2}
\nn
\\
&& = \,-\frac{\la}{32\pi^2} \bigg\{
\int_0^\Om dp^2
- m^2 \int_0^\Om \frac{p^2 dp^2}{p^2+m^2}\bigg\}
\,=\, -\,\frac{\la}{32\pi^2} \bigg\{ \Om^2
- m^2 \ln{\frac{\Om^2}{m^2}}\bigg\},
\label{cutoff}
\eeq
where the $O \big(\Om^{-1} \big)$-terms were omitted as irrelevant.

The dimensional regularization of this diagram is not equally simple,
but will prove instructive for the future. We have
\beq
I_4  &\,\, \longrightarrow \,\,& I_{2\om}
= \int \frac{d^{2\om}p}{(2\pi)^{2\om}} \cdot \frac{1}{p^2+m^2}
= \frac{2\om \cdot \pi^\om}{(2\pi)^{2\om}\Ga{(\om+1)}} \int_0^\infty \frac{p^{2\om-1}dp}{p^2+m^2}\,\,.
\label{I2w}
\eeq
Remember that $\Ga{(\om+1)} = \om\Ga{(\om)}$, hence
\beq
I_{2\om} &=& \frac{2\pi^\om}{(2\pi)^{2\om}}
\cdot \frac{1}{\Ga{(\om)}}
\int_0^\infty \frac{p^{2\om-1}dp}{p^2+m^2}.
\label{Iom}
\eeq

This integral can be expressed via the Beta-function
\beq
B(x,y) &=& \frac{\Ga{(x)}\Ga{(y)}}{\Ga{(x+y)}}
= \int_0^\infty dt \; t^{x-1}(1+t)^{-x-y}.
\label{B}
\eeq
In (\ref{Iom}), we denote $p^2=tm^2$ and obtain
\beq
p^{2\om-1}dp = \frac{1}{2}m^{2\om}t^{\om-1}dt,
&& p^2+m^2=m^2(1+t).
\nn
\eeq
Then
\beq
I_{2\om}
&=&
\frac{\pi^\om\,(m^2)^{\om-1}}{(4\pi^2)^\om\,\Ga{(\om)}} \;
\, \int_0^\infty dt \; t^{\om-1} (1+t)^{-1}
\,=\,
\frac{1}{(4\pi)^\om} \;
\frac{(m^2)^{\om-1}}{\Ga{(\om)}} \cdot B(\om,1-\om)
\nn
\\
&=&
\frac{(m^2)^{\om-1}}{(4\pi)^\om} \;
\frac{\Ga{(\om)\Ga{(1-\om)}}}{\Ga{(\om)\Ga{(1)}}}
\,=\, \frac{(m^2)^{\om-1}}{(4\pi)^\om} \; \Ga{(1-\om)},
\label{I2}
\eeq
where we identified $x-1=\om-1$ and $-x-y=-1$
as arguments of (\ref{B}).

One can use eq. (\ref{gamma17}) to rewrite the result (\ref{I2})
\beq
I_{2\om}
&=&
\frac{(m^2)^{\om-1}}{(4\pi)^\om}
\Big(-\frac{1}{2-\om}+\gamma-1\Big)
\nn
\\
&=& \frac{m^2}{(4\pi)^2}(\mu^2)^{\om-2}
\Big(\frac{m^2}{4\pi\mu^2}\Big)^{\om-2}
\Big(-\frac{1}{2-\om}+\gamma-1\Big),
\eeq
where $\mu$ is a {\it renormalization parameter}, with the
mass dimension, \ $[\mu]=[m]$.
Furthermore,
\beq
\Big(\frac{m^2}{4\pi\mu^2}\Big)^{\om-2}
&=& e^{(\om-2) \ln \big(\frac{m^2}{4\pi\mu^2}\big)}
= 1 + (2-\om) \ln{\Big(\frac{4\pi\mu^2}{m^2}\Big)} + \cdots\,,
\label{12}
\eeq
and we finally arrive at
\beq
I_{2\om}
&=&
\frac{m^2}{(4\pi)^2}(\mu^2)^{\om-2}
\Big[-\frac{1}{2-\om}+\gamma-1
-\ln{\Big(\frac{4\pi\mu^2}{m^2}\Big)}\Big]
\nn
\\
&=&
m^2(\mu^2)^{\om-2}\Big[\frac{2}{\varepsilon}
+\frac{\gamma}{(4\pi)^2}
-\frac{1}{(4\pi)^2}
-\frac{1}{(4\pi)^2}\ln{\Big(\frac{4\pi\mu^2}{m^2}\Big)}\Big],
\eeq
which leads us to
\beq
\frac{1}{2}\,I_{go}= -\lambda m^2(\mu^2)^{\om-2}\Big[\frac{1}{\varepsilon}
+\frac{\gamma}{2(4\pi)^2}
-\frac{1}{2(4\pi)^2}
-\frac{1}{2(4\pi)^2}\ln{\Big(\frac{4\pi\mu^2}{m^2}\Big)}\Big].
\label{f}
\eeq
The next observation is that one can always redefine $\mu$ and absorb
the term $\ga-1$ into the $\ln{\mu}$. Of course, this is not a
compulsory operation.

The comparison between (\ref{f}) and the result in the cut-off
regularization (\ref{cutoff}) shows that in dimensional regularization
there is nothing like the quadratic divergences ${\cal O}(\Om^2)$.
On the other hand, there is a direct relationship between the leading
logarithm $\ln{\frac{\Om}{m}}$ term and 	$\frac{1}{\vp}$ - term.
The correspondence is given by the relation
\beq
\ln{\frac{\Om^2}{m^2}}
\,\, &\longleftrightarrow& \,\,
-\, \frac{\mu^{n-4}}{\vp},
\qquad
\vp = (4\pi)^2 (n-4),
\label{ln}
\eeq
which is universal and holds for all logarithmically divergent
diagrams. Let us note that this is a particular manifestation of the
general rule. The leading logarithms are the same in {\it all}
regularization schemes \cite{Salam51}.

The expression (\ref{f}) has no dependence on the external
momenta and therefore does not contribute to the nonlocal part.
However, the situation is different for other loop integrals.

\subsubsection{UV divergence and the nonlocal form factor}
\label{sect4}

As a second example, consider the diagram
\beq
\includegraphics[keepaspectratio=true,width=4.0cm,height=2.6cm]
{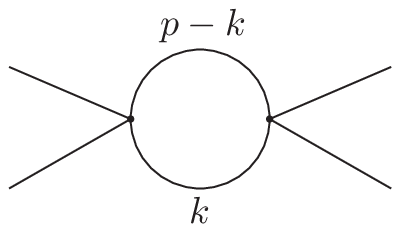}
&=& \frac{\la^2}{2} \int \frac{d^4p}{(2\pi)^4} \cdot
\frac{1}{(p^2+m^2)[(p-k)^2 +m^2]}.
\label{xy}
\eeq

As a first step, we derive the divergent part of (\ref{xy}) in
the cut-off regularization. For this end we make the following
transformation:
\beq
I_4 &=& \int \frac{d^4p}{(2\pi)^4} \;
\frac{1}{(p^2+m^2)[(p-k)^2 +m^2]}
\nn
\\
&=& \int \frac{d^4p}{(2\pi)^4} \; \frac{1}{(p^2+m^2)^2}
+ \int \frac{d^4p}{(2\pi)^4} \; \frac{1}{p^2+m^2} \;
\Big[ \frac{1}{[(p-k)^2 +m^2]} - \frac{1}{p+m^2}\Big].
\mbox{\qquad}
\eeq
Remember that $\,d^4p = \pi^2 p^2dp^2$ in $n=4$. Therefore,
the first integral is logarithmically divergent and the second one
is finite. Then
\beq
I_4^{div}
&=&
\int_0^{\Om} \frac{p^2dp^2}{(4\pi)^2(p^2+m^2)}
= \frac{1}{(4\pi)^2} \ln{\frac{\Om^2}{m^2}} + (\text{finite terms})
\nn
\eeq
and hence (\ref{xy}) is
\beq
\frac{\la^2}{2(4\pi)^2} \ln{\frac{\Om^2}{m^2}}
\,\,+\,\, (\text{finite terms}).
\label{f2}
\eeq

Let us now start with the dimensional regularization calculation.
First we have to define
\beq
I_{2\om} &=& \int \frac{d^{2\om}p}{(2\pi)^{2\om}} \cdot
\frac{1}{(p^2+m^2)[(p-k)^2+m^2]}.
\label{I3}
\eeq
Obviously, at $\om=2$ the integral $I_{2\om}$ coincides with $I_4$, and
also $I_{2\om}$ is analytic on a complex plane in the vicinity of $\om=2$,
where it has a pole (as we will see in brief).

We can use Feynman formula (simplest version)
\beq
\frac{1}{ab} &=& \int_0^1 \frac{d\al}{[a\al +b(1-\al)]^2}.
\label{ab}
\eeq
Using (\ref{ab}), one can cast (\ref{I3}) into the form
\beq
I_{2\om} &=& \int_0^1 d\al \int \frac{d^{2\om}p}{(2\pi)^{2\om}} \;
\frac{1}{[(p-\al k)^2 +a^2]^2},
\label{I4}
\\
\mbox{where} &&
\quad a^2=m^2+\al(1-\al)k^2 .
\label{I4a}
\eeq

Since (\ref{I4}) is convergent on the complex plane, one can make a
shift of the integration variable, $\,p_\mu \to p_\mu -\al k_\mu$. A
simple calculation gives us
\beq
I_{2\om} &=& \int_0^1 d\al \int \frac{d^{2\om}p}{(2\pi)^{2\om}} \;
\frac{1}{(p^2+a^2)^2} \; .
\label{I5}
\eeq

The main advantage of (\ref{I5}) is that is does not depend on the
angles. One can use the same steps that took us from  (\ref{I2w}) to
(\ref{Iom}), and the change of variable $\,p^2 = a^2t$, to arrive at
\beq
I_{2\om}
&=&
\int_0^1 d\al \int \frac{2\pi^\om}{(2\pi)^{2\om} \Ga{(\om)}}
\,dp \, p^{2\om-1} (p^2 + a^2)^{-2}
\nn
\\
&=&
\int_0^1 d\al \int_0^\infty \frac{dt}{(4\pi)^\om \Ga{(\om)}} \;
a^{2\om -4} \; t^{\om -1} (1+t)^{-2}.
\eeq
Comparing this to (\ref{B}), we identify $x=\om$ and $y=2-\om$. Then
\beq
I_{2\om} = \frac{1}{(4\pi)^\om} \int_0^1 d\al \;
\frac{\Ga{(\om)}\Ga{(2-\om)}}{\Ga{(\om)}\Ga{(2)}} \; a^{2\om-4}.
\nn
\eeq
Remember that $\Ga{(2)} = 1$ and $a^2 = m^2 + \al(1-\al)k^2 $, while
$\Ga{(2-\om)} = \frac{1}{2-\om} - \gamma $.
Then
\beq
I_{2\om}
&=&
\frac{1}{(4\pi)^\om} \Big( \frac{1}{2-\om} -\ga \Big)
\int_0^1 d\al \,\big[m^2 + \al(1-\al) k^2\big]^{\om-2}.
\label{Gint}
\eeq
Let us denote $\ta = \frac{k^2}{m^2}$ and transform
\beq
&&
\big[
m^2 + \al(1-\al) k^2\big]^{\om-2}
\,=\,
(m^2)^{\om-2} \; e^{(\om-2) \ln \big[1 + \al(1-\al)\ta\big]}
\nn
\\
&=&
(m^2)^{\om-2} \big[1 - (2-\om) \ln \{1+\al(1-\al)\ta\}
\,+\, O(\om-2)^2\big].
\label{tau}
\eeq
Replacing this expression into (\ref{Gint}), we arrive at
\beq
I_{2\om} &=& \frac{1}{(4\pi)^\om}
\Big( \frac{1}{2-\om} - \gamma \Big)
(m^2)^{\om-2} \Big[ 1 - (2-\om) \int_0^1 d\al \ln{[1 + \al (1-\al)\ta]} \Big]
\nn
\\
&=& \frac{(m^2)^{\om-2}}{(4\pi)^\om} \Big[ \frac{1}{2-\om} - \gamma
- \int_0^1 d\al \,\ln\{1 + \al(1-\al)\ta\} \Big].
\label{I6}
\eeq

In the last formula, the first term is the divergence and the integral
over $\al$ represents the nonlocal form factor, which is the desired
physical result. Also,
\beq
\frac{(m^2)^{\om-2}}{(4\pi)^\om} &=& \frac{(\mu^2)^{\om-2}}{(4\pi)^2}
\Big( \frac{m^2}{4\pi\mu^2} \Big)^{\om-2}
= \frac{(\mu^2)^{\om-2}}{(4\pi)^2} \; e^{(2-\om) \ln{ \Big( \frac{4\pi\mu^2}{m^2}}\Big) }
\nn
\\
&=& \frac{(\mu^2)^{\om-2}}{(4\pi)^2} \Big[ 1 + (2-\om)
\ln{\Big( \frac{4\pi\mu^2}{m^2} \Big)} + O(2-\om)^2 \Big].
\label{m}
\eeq
An elementary (albeit deserving to be checked by the reader)
integration provides
\beq
&&
\int_0^1 d\al \; \ln\{1 + \al(1-\al)\ta\} \,=\, -2Y ,
\label{int}
\\
\mbox{where}
\qquad
a^2 &=& \frac{4\ta}{\ta+4} = \frac{4k^2}{k^2+4m^2}
\qquad
\mbox{and}
\qquad
Y \,=\, 1 - \frac{1}{a}\ln{\Big| \frac{2+a}{2-a} \Big|}.
\label{Y}  
\eeq
Replacing (\ref{m}) and (\ref{int}) into (\ref{I6}), we arrive at
\beq
I_{2\om} &=& \frac{(\mu^2)^{\om-2}}{(4\pi)^2} \Big[ \frac{1}{2-\om}
- \gamma + \ln{\Big( \frac{4\pi\mu^2}{m^2} \Big)} + 2Y \Big]
\label{I7}
\\
&=& (\mu^2)^{\om-2} \Big[ - \frac{2}{\varepsilon}
- \frac{\ga}{(4\pi)^2}
+ \frac{1}{(4\pi)^2} \ln{\Big(\frac{4\pi\mu^2}{m^2} \Big)}
+ \frac{2Y}{(4\pi)^2} \Big].
\nn
\eeq

The last thing to do is to explore the form factor Y in the two
extremes, namely high- and low-energy limits,
\beq
&&
\mbox{1) \ \ \  UV, }
\quad k^2 \gg m^2, \quad \mbox{that means} \quad \ta \gg 1,
\nn
 \\
&&
\mbox{2) \ \ \  IR,}
\ \ \quad  k^2 \ll m^2, \quad \mbox{that means} \quad  \ta \ll 1.
\eeq

\textbf{1)} \ \ Consider the UV regime, that means $\,k^2 \gg m^2\,$
and $\,\ta \gg 1$. \ Thus,
\beq
a^2 = \frac{4k^2}{k^2+4m^2} = \frac{4}{1+\frac{4m^2}{k^2}}
= 4 \Big(1 - \frac{4m^2}{k^2} + \cdots \Big),
\nn
\eeq
hence $a \approx 2 - \frac{4m^2}{k^2}$. 
Then $2+a \approx 4-\frac{4m^2}{k^2}$ and
$2-a \approx \frac{4m^2}{k^2}$, such that
\beq
Y & \cong & 1 - \frac{1}{2} \ln{\Big(\frac{k^2}{m^2}\Big)}.
\label{Y_UV}
\eeq
In this case, $I_{2\om}$ in (\ref{I7}) includes the combination
\beq
I_{2\om} &=& \frac{(\mu^2)^{\om-2}}{(4\pi)^2} \Big[ \frac{1}{2-\om}
+ \ln{\Big( \frac{4\pi\mu^2}{m^2} \Big)} + 2 - \gamma
- \ln{\Big( \frac{k^2}{m^2} \Big)} \Big]
\nn
\\
&=& \frac{(\mu^2)^{\om-2}}{(4\pi)^2} \Big[ - \,\frac{2}{n-4}
\,+\, \ln{\Big( \frac{\mu^2}{k^2} \Big)} + \text{constant} \Big] .
\label{IomUV}
\eeq
Let us stress that this is a very significant and important relation,
as it shows two things at once. The first point is that the
large-$k^2$ limit means large $\mu^2$ limit and v.v. Thus, it
is sufficient to establish the large $\mu$ limit within the $MS$
scheme of renormalization, to know the physical UV limit, that
is the behavior of the quantum system at high energies. Let us
remember that we already know well how to explore the
large-$\mu$ limit throughout the usual renormalization group,
including in curved space \cite{nelspan82,tmf,book}.

The second aspect is that one can always restore large-$\mu^2$
limit from the coefficient of the divergent term with the
$\frac{1}{\vp}$-factor. On the other hand, there is also a direct
correspondence with the large cut-of limit in (\ref{f2}). All in all,
we can say that the UV limit is pretty well controlled by the leading
logarithmic divergences, which can be derived easily within the
heat-kernel methods, even without the use of Feynman diagrams.
\vskip 2mm

\textbf{2)} \ \ Consider now the IR regime, when $k^2 \ll m^2$,
or, equivalently, $\ta \ll 1$. Then
\beq
a^2 \,\, \sim \,\, \frac{k^2}{m^2} \ll 1
\nn
\eeq
and hence $\,a \sim \frac{k}{m}$. As a consequence of this,
\beq
\ln{ \frac{2+a}{2-a}}
\,\,\approx\,\, \ln{\frac{2+\frac{k}{m}}{2-\frac{k}{m}}}
\,\,\approx\,\, \ln\Big(1+\frac{k}{m}\Big)
\,\,\approx\,\, \frac{k}{m},
\nn
\eeq
and therefore
\beq
Y &=& 1 - \frac{1}{a} \ln{\Big| \frac{2+a}{2-a} \Big|}
\,\,\approx\,\, 1 - \frac{m}{k} \; \frac{k}{m}
\,\,\approx\,\, 0.
\eeq
In the zero-order approximation there is no non-local form factor in
the IR. This means there is no \ $\mbox{ln}{\Big(\frac{k^2}{m^2}\Big)}$
that corresponds to the divergences $\frac{1}{\varepsilon}$, or
$\ln{\Om}$. The next orders of expansion read
\beq
Y &=& -\frac{1}{12}
 \; \frac{k^2}{m^2}
+ \frac{1}{120} \Big( \frac{k^2}{m^2} \Big)^2 + \cdots .
\label{IR-Y}
\eeq
The first term is the evidence that the decoupling is quadratic in
this case. The same quadratic dependence takes place in all cases
when we can check it. In the IR limit, the divergences and momentum
dependence do not correlate with
each other. This phenomenon is called IR decoupling, or ``decoupling
theorem''. It was discovered in QED in 1975 by Appelquist and
Corrazzone \cite{AC}.

As we already mentioned above, for the tadpole diagram (\ref{f})
there is no any non-local form-factor. In this case, one may think
that the UV divergence is ``artificial'', but this is not a correct
viewpoint, because in general the logarithmic form factor corresponds
to the {\it sum} of all divergent contributions, including the ones of
the tadpoles. This argument is necessary for the correct
calculation in curved space-time \cite{apco}, since only taking it
into account one can establish the relationship between leading logs of
momentum in the UV and the dependence on $\mu$. Now we can
better understand this point, since we know that the
$\mu$-dependence in (\ref{f}) appears together and in not separable
from the divergent term.

\section{Non-local form factors in curved space time}
\label{nonloca}

Let us now turn around to the functional method and consider in
full detail the derivation of the form factors using the heat-kernel
solution of \cite{bavi90}.

The one-loop contribution to the Euclidian Effective Action for a
massive field is defined as the trace of the coincidence limit of the
logarithm of determinant of the bilinear form of the action or,
equivalently,  as an integral of the heat-kernel over the proper
time $s$,
\beq
\label{1}
{\overline{\Gamma}}^{(1)} &=& \frac{1}{2}\,\Tr\ln
 \left( - {\hat 1}\cx  + m^2 - \hat{P}
 + \frac{{\hat 1}}{6}\,R \right)
 \,\,=\,\, \frac{1}{2}\int_0^\infty{\frac{ds}{s}}\,\Tr K(s)\,.
\eeq
This formula is valid for bosonic fields in Euclidian space-time, while
for fermions, the overall sign in the equation (\ref{1}) has to be changed.
Here $K(s)$ is the heat-kernel of the bilinear form of the classical
action of the theory, $\cx = \na^2$ is the covariant Laplacian, and
\beq
\Tr K(s)
&=& \frac{(\mu^2)^{2-\om}}{(4\pi s)^{\omega}}
\int d^4x \sqrt{g}\ e^{-sm^2}\,\tr \big\{{\hat 1}
+ s\hat{P} + s^2 \big[ {\hat 1}\ R_{\mu\nu}f_1(\tau)R^{\mu\nu}
\label{2}
\\
&+&
{\hat 1} R f_2(\tau)R
+ \hat{P}f_3(\tau)R + \hat{P}f_4(\tau)\hat{P}
+ \hat{\cal R}_{\mu\nu}f_5(\tau)\hat{\cal R}^{\mu\nu} \big]\big\}.
\nonumber
\eeq
Here $\tau=-s\Box$ and we use notation
$\hat{\cal R}_{\mu\nu} = [\nabla_\mu, \nabla_\nu]$ from \cite{bavi85}.
Let us note that we use $\Box = g^{\mu\nu}\na_\mu\na_\nu$ even in
Euclidean space. In equation (\ref{2}), the terms between braces are
matrices in the space of the fields (scalar, vector, or fermion). The
zero-order term proportional to $\,\tr {\hat 1}\,$  corresponds to
quartic divergence, or to the coefficient $a_0$ in the
Schwinger-DeWitt expansion. The term with $\,s\tr \hat{P}$
corresponds to the quadratic divergences, or to the $a_1$ coefficient,
and all the rest corresponds to the logarithmic divergences  and is
related to $a_2$-coefficient plus finite terms. The infinite tower of
the terms of third- and higher-orders in curvatures are omitted in
this formula.

As we already saw in the diagrams-based approach, $a_0$- and
$a_1$-terms can be eliminated by the choice of regularization
scheme, so we shall mainly focus on $a_2$ and related
terms\footnote{Indeed, there are finite nonlocal surface terms
related to $a_1$, and those are regularization-independent. One
can learn about this aspect in \cite{omar2013,sebastian}.}.
The functions $f_{1...5}$ have the form \cite{bavi90}
\beq
&&
f_1{(\tau)} = \frac{f{(\tau)} - 1 + {\tau}/6}{{\tau}^2}\, ,
\qquad\,\,
f_2{(\tau)} = \frac{f{(\tau)}}{288}
+ \frac{f{(\tau)} - 1}{24{\tau}}-\frac{f(\tau)-1+\tau /6}{8\tau^2}\,  ,
\nn
\\
&&
f_3{(\tau)} =  \frac{f{(\tau)}}{12} + \frac{f{(\tau)} - 1}{2{\tau}}\, ,
\qquad\,
f_4{(\tau)} =  \frac{f{(\tau)}}{2}\, ,
\qquad\,
f_5{(\tau)} =  \frac{1 - f{(\tau)}}{2{\tau}}\, ,
\label{4}
\eeq
where
\begin{equation}\label{5}
f{(\tau)} = \int_0^1 {d{\alpha} \ e^{-{\alpha}(1-{\alpha}){\tau}}}
\qquad
\mbox{and}
\qquad
\tau = -s \Box \, .
\end{equation}
In what follows we shall describe the derivation of the integral in
(\ref{1}) for a particular case of a massive scalar field.

\section{Form factors for the massive scalar theory}
\label{nlscal}

The action for the theory with the general non-minimal coupling
to the scalar curvature is
\beq
S &=& \int d^4 x \sqrt{g}
\ \left\{\frac{1}{2} g^{\mu\nu}{\pa_{\mu}}{\ph} \ {\pa_{\nu}{\ph}}
+ \frac{1}{2}(m^2 + {\xi}R){\varphi}^2\right\}
\label{1s}
\eeq
and therefore
\beq
\hat{P} &=&
- \left(\xi - \frac{1}{6}\right)R
\qquad \mbox{and} \qquad
\hat{\cal R}_{\al\be} = 0\,,
\eeq
In the case under consideration ${\hat 1}=1$, also it is good to
note that we did not include $m^2$ into $\hat{P}$, as it was done
with ${\cal P}$ in (\ref{P}).

According to (\ref{1}) and (\ref{2}), the bilinear in curvatures part
of the effective action can be given by the proper-time integral of
the heat kernel,
\beq
\nonumber
{\bar \Ga}^{(1)}
&=&
\frac{1}{2}\int_0^\infty{\frac{ds}{s}}\,
\frac{\mu^{2(2-\omega)}}{(4\pi s)^\om}
\int d^{4}x \sqrt{g}\,e^{-sm^2}\,\tr
\Big\{
1 + s\hat{P}
+ s^2 \big[ R_{\mu\nu}f_1(-s{\nabla}^2)R^{\mu\nu}
\\
&+&
R f_2(-s{\nabla}^2)R
+ \hat{P}f_3(-s{\nabla}^2)R
+ \hat{P}f_4(-s{\nabla}^2)\hat{P}
+ \hat{\cal R}_{\mu\nu}f_5(-s{\nabla}^2)\hat{\cal R}^{\mu\nu} \big]
\Big\}\,.
\label{6}
\eeq

Let us derive the integrals over proper time in Eq.~(\ref{6}), starting
from the simplest ones.

\subsection{Zero-order term}

Consider the term which corresponds to the $a_0$-coefficient in
the expression for the divergences,
\beq
\label{7}
{\bar \Ga}^{(1)}_{0}
&=&
\frac{1}{2}\int_0^\infty{\frac{ds}{s}}\,
\frac{\mu^{2(2-\om)}}{(4\pi s)^{\om}}
\int d^4x  \sqrt{g}\,e^{-sm^2} \,.
\eeq
It proves useful making a change of variables as follows
\beq
\label{mv}
s = \frac{t}{m^2}\,,
\quad
ds = \frac{dt}{m^2}\,,
\quad
\frac{ds}{s^{1+\om}} \,=\, \frac{dt \,m^{2\om}}{t^{1+\om}}\,.
\eeq
Then the integral becomes
\beq
\label{9}
\nonumber
{\bar \Ga}^{(1)}_{0}
&=&
\frac{1}{2}\int d^4x \ \sqrt{g} \,\frac{\mu^{2(2-\om)}}{(4\pi)^\om} \,m^{2\omega}\int_0^\infty{\frac{dt}{t^{1+\omega}} \ e^{-t}}
\\
\nonumber
&=&
\frac{1}{2}\int d^4x \sqrt{g}
\ \frac{m^4}{2(4\pi)^2}
\left(\frac{m^2}{4{\pi}{\mu}^2}\right)^{\om - 2}
\Big[\frac{1}{(2-{\omega})}+\frac{3}{2} + O(2-\omega) \Big]
\\
&=&
\frac{1}{{2(4\pi)^2}}
\int d^4x \sqrt{g} \,\Big[\frac{1}{2-\omega}
+ \ln \Big(\frac{4{\pi}{\mu}^2}{m^2}\Big)
+ \frac{3}{2}\Big]\,\frac{m^4}{2}\,.
\eeq

In the calculation presented above we have used the relations
(\ref{Ga3}) and (\ref{12}).
It proves useful to introduce the following new notation:
\beq
\frac{1}{\vp_{\om,\mu}}
&=&
\frac{1}{2(4\pi)^2}\,
\Big[\frac{1}{\om-2} - \ln \Big(\frac{4\pi\mu^2}{m^2}\Big)\Big]\,.
\label{epsi}
\eeq
Then the one-loop contribution to the cosmological constant term
(\ref{9}) becomes
\beq
{\bar \Ga}^{(1)}_{0}
&=&
\int d^4x \sqrt{g} \,
\bigg[
\,-\,\frac{1}{\vp_{\om,\mu}}
\,+\, \frac{3}{4\,(4\pi)^2}\bigg]\,\frac{m^4}{2}\,.
\label{9eps}
\eeq
We can observe that this expression consists of the UV divergence,
corresponding $\,\rm{ln}\mu$-term hidden in $1/\vp_{\om,\mu}\,$ and
the irrelevant constant term, which can be easily absorbed into
$1/\vp_{\om,\mu}$ by changing $\mu$. There is no non-local
form factor in the expression (\ref{9eps}). As we already explained
above,  this is a natural result since such a form factor should be
constructed from $\cx$, which, when acting on $m^4$, gives zero.

\subsection{First-order term}

In the next order in $s$, we meet
\beq
\label{14}
\nonumber
{\bar \Ga}^{(1)}_{1}
&=&
 \frac{1}{2}\int_0^\infty{\frac{ds}{s}}
 \frac{\mu^{2(2-\om)}}{(4\pi s)^{\omega}}\int d^4x \sqrt{g}
 \ e^{-sm^2}\,\tr (s\hat{P})
\\
&=&
-\,\frac{1}{2}\int_0^\infty \frac{ds}{s^\om}\,e^{-sm^2}
\int d^4x \sqrt{g}\,\frac{\mu^{2(2-\om)}}{(4\pi)^{\om}}
\,\Big(\xi - \frac16\Big) R
\nonumber
\\
&=&
-\,\frac{1}{2}\,
\frac{\mu^{2(2-\om)}}{(4\pi)^{\om}}
\,m^{2(\omega-1)}\int d^4x \sqrt{g} \Big(\xi - \frac16\Big)
\,\Ga(1-\omega)\, R
\nonumber
\\
&=&
\bigg[-\,\frac{1}{\vp_{\om,\mu}} + \frac{1}{2{(4\pi)^{2}}}
\bigg]
\,\Big(\xi - \frac16\Big) \int d^4x \sqrt{g}\,m^2R\,.
\eeq
where we used the expansion (\ref{12}), definition (\ref{epsi})
and the relation
\beq
\tr \hat{P} &=& -\,\Big(\xi - \frac{1}{6}\Big) R .
\eeq

Thus, without invoking the surface terms \cite{sebastian}, the
effective action is local and the logarithmic dependence on the
renormalization parameter $\mu$ is completely controlled by
the pole of $\frac{1}{2-\om}$. The
results (\ref{9eps}) and (\ref{14}) enable one to construct
the Minimal Subtraction - based renormalization group equations
for the cosmological constant density and Newton constant, but
they do not provide the nonlocal terms hidden behind the
renormalization group.

\subsection{Second-order terms}

Working with the next-order terms is much more involved.
We shall calculate them one by one, to find the coefficients
$l^*_{1...5}$ and $l_{1...5}$, which define the final form factors
of the $R_{\mu\nu}\cdot R^{\mu\nu}$- and $R\cdot R$-terms.
The general expression in the second order in curvature can be
cast into the form
\beq
\label{18}
\nonumber
{\bar \Ga}^{(1)}_{2} &=& {\bar \Ga}_{R_{\mu\nu}^2} + {\bar \Ga}_{R^2}
\,=\,\sum_{k=1}^{5}{\bar \Ga}_k\,.
\eeq

Finally, remembering that $\hat{\cal R}_{\al\be} = 0$, we have to evaluate
\beq
\label{24}
\nonumber
{\bar \Ga}^{(1)}_{2}
&=&  \frac{1}{2}
\int_0^\infty{ds  \, e^{-sm^2} \,{s^{1-\omega}}}
\, {\frac{(\mu^2)^{2-\omega}}{(4\pi)^{\omega}}\int d^4x \sqrt{g}}\,
\Big\{R_{\mu\nu}f_1(-s\Box)R^{\mu\nu}
\\
&+& R\, \Big[
f_2 (-s\Box)
- \Big(\xi-\frac16\Big) f_3(-s\Box)
+ \Big(\xi-\frac16\Big)^2 f_4(-s\Box)\Big]R\Big\} \,.
\eeq
By replacing the relations (\ref{4}) and (\ref{5})
into (\ref{18}) and, once again, $-s\Box$ by $\tau$, we arrive at
\beq
\label{25}
{\bar \Ga}^{(1)}_{2}
&=& \frac{1}{2}\int_0^\infty ds\, e^{-sm^2} \,s^{1-\om}\, {\frac{(\mu^2)^{2-\omega}}{(4\pi)^{\omega}}
\int d^4x \sqrt{g}}\,
\Big\{
R_{\mu\nu}
\Big[\frac{f(\tau)}{\tau^2}
- \frac{1}{\tau^2} + \frac{1}{6\tau}\Big]R^{\mu\nu}
\\
&+&
R\,
\Big[
  \Big(\frac{1}{288}
- \frac{{\tilde \xi}}{12}
+ \frac{{\tilde \xi}^2}{2}\,\Big) f(\tau)
+ \Big(\frac{1}{24} - \frac{{\tilde \xi}}{2}\Big)\frac{f(\tau)}{\tau}
- \frac{f(\tau)}{8\tau^2}
+ \Big(\frac{\tilde \xi}{2}
- \frac{1}{16}\Big)\frac{1}{\tau}
+ \frac{1}{8\tau^2}\Big]\,R
\Big\} .
\nonumber
\eeq
where the condensed notation ${\tilde \xi}=\xi-1/6$ has been used.

It proves useful to introduce new set of coefficients,
\beq
&&
l_1^* = 0,
\qquad
l_2^* = 0,
\qquad
l_3^* = 1,
\qquad
l_4^* = \frac{1}{6},
\qquad
l_5^* = -1
\qquad
\mbox{and}
\label{29}
\\
&&
l_1 = \frac{1}{288} - \frac{1}{12}\,{\tilde \xi}
+ \frac{1}{2}\,{\tilde \xi}^2\,,
\quad\,
l_2 = \frac{1}{24} - \frac{1}{2}\,{\tilde \xi}\,,
\quad\,
l_3 = -\frac{1}{8} = - l_5\,,
\quad\,
l_4 = -\frac{1}{16} + \frac{1}{2}\,{\tilde \xi}\,.
\nn
\eeq
Furthermore, the basic integrals from Eq.~(\ref{25})
will be denoted as (remember $\tau = -s\Box$)
\beq
\nonumber
M_1 &=&
\int_0^\infty \frac{ds}{(4\pi)^\om}\,e^{-m^2s}\,s^{1-\om}\,f(\tau)
\,=\,
\frac{\mu^{2(2-\om)}}{(4\pi)^{\om}}
\int_0^\infty dt \, e^{-t} \, t^{1-w}\,{f(tu)}\,,
\\
\nonumber
M_2 &=&
\int_0^\infty \frac{ds}{(4\pi)^\om}\,e^{-m^2s}\,s^{-\om}\,f(\tau)
\,=\,
\frac{\mu^{2(2-\om)}}{(4\pi)^{\om}}
\int_0^\infty dt \, e^{-t} \, \frac{f(tu)}{u\,t^{w}} \,,
\\
\nonumber
M_3 &=&
\int_0^\infty \frac{ds}{(4\pi)^\om}\,e^{-m^2s}\,s^{-1-\om}\,f(\tau)
\,=\,
\frac{\mu^{2(2-\om)}}{(4\pi)^{\om}}
\int_0^\infty dt \, e^{-t}\,\frac{f(tu)}{u^2\,t^{1+w}} \,,
\\
\nonumber
M_4 &=&
\int_0^\infty \frac{ds}{(4\pi)^\om}\,e^{-m^2s}\,s^{-\om}
\,=\,
\frac{\mu^{2(2-\om)}}{(4\pi)^{\om}}
\int_0^\infty dt \, e^{-t} \, \frac{1}{u\,t^{w}} \,,
\\
M_5 &=&
\int_0^\infty \frac{ds}{(4\pi)^\om}\,e^{-m^2s}\,s^{-1-\om}
\,=\,
\frac{\mu^{2(2-\om)}}{(4\pi)^{\om}}\, m^{2(w-2)}
\int_0^\infty dt \, e^{-t} \, \frac{1}{u^2\,t^{1+w}}\,,
\label{33}
\eeq
where we already made change of variables (\ref{mv}) and also
denoted (it is the Fourier conjugate of $\tau$ in Eq.~(\ref{tau}))
\beq
u &=& - \frac{m^2}{\Box}\,.
\label{u}
\eeq

A relevant observation is that all individual features of the given
theory (like the scalar in the present case) are encoded into the
coefficients (\ref{29}), while the integrals (\ref{33}) are universal
in the sense, they will be the same for any theory which provides us
an operator of the form (\ref{Pih}), at least with $\hat{h}^\al=0$.
Thus, the derivation of the same set of integrals  (\ref{33}) enables
one to derive the form factors in many interesting cases.

In the new notations (\ref{29}) and (\ref{33}) the second-order
part of the one-loop effective action can be cast into the form
\beq
\label{32}
{\bar \Ga}^{(1)}_{2} &=& {\bar \Ga}^{(1)}_{R_{\mu\nu}^2}
+ {\bar \Ga}^{(1)}_{R^2}
\,=\,\frac{1}{2}\,
\int d^4x \sqrt{g}\,\sum_{k=1}^5\Big\{
R_{\mu\nu} \, l_k^* M_k \, R^{\mu\nu}
+ R \, l_k M_k \, R\Big\}.
\eeq

Let us now calculate the integrals (\ref{33}). We will need
expansion (\ref{12}) and the previous formulas (\ref{gama2}),
(\ref{Ga3}) for the gamma-functions.

Taking these formulas into account, the derivation of $M_4$ and
$M_5$ is an elementary exercise and we just present the results,
\beq
\label{37}
M_4 &=&
-\,\frac{1}{(4\pi)^2\,u}\,\Big\{1 + \frac{1}{2-\om}
+ \ln \Big(\frac{4\pi\mu^2}{m^2}\Big)\Big\}\,
\\
\label{38}
M_5 &=&
\frac{1}{(4\pi)^2\, 2u^2}\,
\Big\{\frac32 + \frac{1}{2-\om}
+ \ln \Big(\frac{4\pi\mu^2}{m^2}\Big)\Big\} \,.
\eeq

In order to calculate the remaining three integrals, we introduce
a new notations
\beq
a^2 &=& \frac{4u}{u+4} \,=\,
\frac{4 \Box}{\Box - 4m^2} \,,
\quad
\mbox{and}
\quad
\frac{1}{u} \,=\, \frac{1}{a^2} - \frac14 \,,
\label{a}
\eeq
being exactly the Fourier image of (\ref{int}) in the Euclidean space,
with
\beq
a^2 &=& \frac{4 k^2}{k^2 + 4m^2}\,\geq\,0
\,,\quad \mbox{also} \quad
a^2 \leq 4\,.
\label{amom}
\eeq
We can assume, for definiteness, that $a$ changes from $a=0$ in
the IR to $a=2$ in the UV.

Furthermore, we will need the following integral
\beq
A
&=&
-\,\frac{1}{2}\int_0^1 d\al \,\ln [1 + \al(1-\al)u]
\,=\, 1 - \frac{1}{a} \, \ln \left| \frac{2+a}{2-a}\right| \,,
\label{A}
\eeq
which is the same as $Y$ from Eq.~(\ref{Y}) in the
coordinate space.

The remaining calculation of the first three integrals is not
complicated and we just give the final results in terms of
$a$ and $A$,
\beq
\label{34}
M_1
&=&
-\,\frac{2}{\vp_{\om,\mu}} \,+\, \frac{2A}{(4\pi)^2}\,,
\\
\label{35}
M_2
&=&
\Big[ -\,\frac{2}{\vp_{\om,\mu}} + \frac{1}{(4\pi)^2}\Big]\,
\Big(\frac{1}{12} - \frac{1}{a^2}\Big)
\,+\, \frac{1}{(4\pi)^2}\,
\Big\{\frac{1}{18}- \frac{4A}{3a^2}\Big\} \,
\\
&=&
\frac{1}{(4\pi)^2}\,
\Big\{\Big[ 1 +
\frac{1}{2-\om} + \ln \Big(\frac{4\pi\mu^2}{m^2}\Big)\Big]
\Big(\frac{1}{12} - \frac{1}{a^2}\Big) - \frac{4A}{3a^2}
+ \frac{1}{18}\Big\} \,,
\\
\label{36}
M_3
&=&
\frac{1}{(4\pi)^2}\Big\{\Big[\frac{3}{2} + \frac{1}{2-\om}
+ \ln \Big(\frac{4\pi\mu^2}{m^2}\Big)\Big]\Big[\frac{1}{2a^4}
- \frac{1}{12a^2} + \frac{1}{160}\Big]
\nonumber
\\
&+& \frac{8A}{15a^4} - \frac{7}{180a^2} + \frac{1}{400}\Big\} \,.
\eeq

Now one can construct a useful combinations for the scalar case, such as
\beq
\label{39}
\nonumber
M_{R_{\mu\nu}^2}
&=&
l_3^* \, M_3 + l_4^* \, M_4 + l_5^* \, M_5
\,=\,
M_3 + \frac{1}{6}\, M_4 - M_5
\\
&=&
\frac{1}{(4\pi)^2}\Big\{\frac{1}{2-\om} \Big(\frac{1}{60}\Big) + \ln \Big(\frac{4\pi\mu^2}{m^2}\Big)\Big(\frac{1}{60}\Big)
+ \frac{8A}{15a^4} + \frac{2}{45a^2} + \frac{1}{150}
\eeq
and
\beq
\label{40}
\nonumber
M_{R^2}
&=&
l_1 \, M_1 + l_2 \, M_2 + l_3 \, M_3 + l_4 \, M_4 + l_5 \, M_5
\\
\nonumber
&=& \Big(\frac{1}{288} - \frac{1}{12}\,{\tilde \xi}
+ \frac{1}{2}\,{\tilde \xi}^2 \Big) M_1
+ \frac12\,\Big(\frac{1}{12} - {\tilde \xi}\Big) M_2
- \frac{1}{8}\, M_3
- \frac12\,\Big(\frac{1}{8} - {\tilde \xi}\Big) M_4
+ \frac{1}{8}\, M_5
\\
\nonumber
&=&
\frac{1}{(4\pi)^2}\,\Big\{
\Big[\frac{1}{2-\om} + \ln\Big(\frac{4\pi\mu^2}{m^2}\Big)\Big]\,
\Big(\frac{1}{2}\,{\tilde \xi}^2
- \frac{1}{180}\Big)
+ A {\tilde \xi}^2 + \frac{2A}{3a^2}{\tilde \xi}
- \frac{A}{6}\, {\tilde \xi}
- \frac{A}{18a^2}
\\
&+&
  \frac{A}{144} - \frac{A}{15a^4}
- \frac{59}{10800} - \frac{1}{180a^2}
+ \frac{1}{18} {\tilde \xi}
\Big\}\,.
\eeq

Finally, we meet
\beq
\label{41}
{\bar \Ga}^{(1)}_{2}
&=&
\frac{1}{2} \int d^4 x \sqrt{g} \,
\Big\{
R_{\mu\nu}\,M_{R_{\mu\nu}^2}\,R^{\mu\nu}
\,+\,
R\, M_{R^2}\,R \,\Big]\} \,.
\eeq
Let us note that there is a third term, related to the
square of the Riemann tensor. However, for any integer $N$
one can prove, utilizing the Bianchi identities and partial
integrations, that (see, e.g., \cite{highderi})
\beq
\label{GB}
E_{4,N} \,=\,
R_{\mu\nu\al\be}\Box^N R^{\mu\nu\al\be}
-4R_{\mu\nu}\Box^N R^{\mu\nu}
+ R\Box^N R \,=\,{\cal O}(R_{\dots}^3)
\,+\,\mbox{total derivatives}.
\eeq
This means that in the bilinear in curvature approximation,
such as the one we discuss here, one can safely use the
reduction formula related to the Gauss-Bonnet term,
\beq
\label{redu}
R_{\mu\nu\al\be} \,f(\Box)\, R^{\mu\nu\al\be}
&=&
4R_{\mu\nu}\,f(\Box)\, R^{\mu\nu} \,-\,R\,f(\Box)\, R\,.
\eeq
As a result, in the curvature-squared approximation, there is no way
to see the non-localities associated to the Gauss-Bonnet combination.
Hence, we can use either $R_{\mu\nu}^2$- and $R^2$-terms, or some
other equivalent basis. For various applications, the most useful
basis consists of the square of the Weyl tensor instead of the square
of the Ricci tensor. The transition can be done using the formulas
\beq
\label{42}
C^2
&=&
R_{\mu\nu\al\be}^2 - 2R_{\mu\nu}^2 + \frac{1}{3} R^2 = E_4 + 2W,
\\
\nonumber
\mbox{where} \quad
W &=& R_{\mu\nu}^2 - \frac{1}{3} R^2 ,
\quad
E_4 \,\,
\mbox{is irrelevant, and}
\\
\tilde{M}_{R^2} &=& M_{R^2} + \frac13\,M_{R_{\mu\nu}^2} .
\label{42m}
\eeq

Introducing the form factors $k_W$ and $k_R$,
\beq
\label{45W}
k_W
&=&
k_{R_{\mu\nu}^2} = \frac{8A}{15a^4} + \frac{2}{45a^2} + \frac{1}{150} \,,
\\
\label{45R}
k_R
&=&
A{\tilde \xi}^2
+ \Big(\frac{2A}{3a^2} + \frac{1}{18} - \frac{A}{6}\Big)\,{\tilde \xi}
- \frac{A}{18a^2} + \frac{A}{144} + \frac{A}{9a^4} - \frac{7}{2160}
- \frac{1}{108a^2}
\eeq
and taking zero-, first- and second-order terms together, one can write
down the effective action up to the second order in curvatures,
\beq
\label{46}
\nonumber
\overline{\Gamma}^{(1)}_{scalar}
&=&
\frac{1}{2(4\pi)^2} \int d^4 x \sqrt{g}\,
\bigg\{\frac{m^4}{2}\Big[\frac{1}{2-\om}
+ \ln \Big(\frac{4\pi\mu^2}{m^2}\Big) + \frac{3}{2}\Big]
\\
\nonumber
&+& \Big(\xi - \frac{1}{6}\Big) m^2 R \Big[\frac{1}{2-\om}
+ \ln \Big(\frac{4\pi\mu^2}{m^2}\Big) + 1\Big]
\\
\nonumber
&+&
\frac12\, C_{\mu\nu\al\be}
\Big[\frac{1}{60 (2-\om)} + \frac{1}{60}
\ln \Big(\frac{4\pi\mu^2}{m^2}\Big)
 + k_W  \Big] C^{\mu\nu\al\be}
\\
&+& R \,
\Big[\,\frac{1}{2} \Big(\xi - \frac{1}{6}\Big)^2 \Big(\frac{1}{2-\om}
+ \ln \Big(\frac{4\pi\mu^2}{m^2}\Big)\Big) + k_R \, \Big] R\bigg\} \, .
\eeq

Let us make some observations concerning the final result for the
vacuum effective action for the scalar field (\ref{46}), with the
form factors (\ref{45W}) and (\ref{45R}). First of all, one has to
stress that this action is essentially non-local in the higher
derivative sector. The result is exact in the derivatives of the
curvature tensor but it is only of the second-order in the curvatures
themselves. On the other hand, the lower-derivative terms, namely
quantum corrections to the cosmological constant and to the term
linear in curvature, do not have non-local parts.

The non-localities derived from the heat-kernel and from the
Feynman diagrams in dimensional regularization, are the same.
This is confirmed by the correspondence between the quantities
$Y$ and $A$ in our two considerations, and also by the original
calculations of \cite{apco} and \cite{omar2013} of the
gravitational form factors.

For the massless (or UV, for the massive field) limit we need to
assume $-\cx/m^2 \gg 1$, in the sense $k^2/m^2 \gg 1$ for the
Euclidean momentum $k$. Then the form factors $k_W$ and $k_R$
can be elaborated following the method of Eq.~(\ref{IomUV}).
According to (\ref{a}), in this limit $a \to 2$. Then,
\beq
k_W &\sim & -\,\frac{1}{120}\,\ln\Big(\frac{- \cx}{\mu^2}\Big)
\,+\,
\mbox{constant and vanishing terms.}
\label{kWinUV}
\\
k_R &\sim & -\,
\frac{1}{2} \Big(\xi - \frac{1}{6}\Big)^2
\,\ln\Big(\frac{ -\cx}{\mu^2}\Big)
\,+\,
\mbox{constant and vanishing terms.}
\label{kR-UV}
\eeq
These relations show that in the UV limit one can restore nonlocal
terms from the logarithmic divergences. On the other hand, for the
massive models out of the UV limit, the nonlocal terms have complex
structure and there is no way to restore them from divergences.

We can
say that the logarithmic UV divergence controls the minimal
subtraction - scheme based renormalization group, covered by the
$\mu$-dependence, and also agrees with the physical behavior of
the theory in UV, that means the logarithmic dependence on the
momenta $p$ in the regime when $(p/m) \to \infty$. The final
observation over the form factor (\ref{45W}) is that the
expression (\ref{kWinUV}) enables one to find the Weyl-squared
part of the conformal anomaly in the massless limit. For this
end one has to use the conformal parametrization of the metric
$g_{\mu\nu}= g^\prime_{\mu\nu}\,\exp\{2\si(x)\}$ and note
that
\beq
\Box &=& e^{-2\si(x)}\big[ \Box^\prime + {\cal O}(\si) \big]\,.
\label{Box}
\eeq
Now, deriving the anomaly by the prescription
\beq
\langle T_\mu^\mu \rangle &=&
- \frac{2}{\sqrt{-g}}\,g_{\mu\nu}
\frac{\de\,\Ga[g_{\mu\nu}]}{\de\, g_{\mu\nu}}
= - \frac{1}{\sqrt{-{\bar g}}}\,e^{- 4\si}
\left.\frac{\de\,\Ga[{\bar g}_{\mu\nu}\,e^{2\si}]}{\de \si}
\,\right|_{{\bar g_{\mu\nu}}\rightarrow g_{\mu\nu},
\si\rightarrow 0}
\label{deriv}
\eeq
we can immediately recover from (\ref{Box}) the $C^2$-part
of the anomaly with the correct coefficient, identical to the
one of the corresponding divergence \cite{PoImpo}.

Let us make one more observation concerning the form factor 
 for the Weyl-squared term. The calculations 
 described above have been done in the Euclidean signature. 
However, if performing the derivation with the Minkowski 
signature and $\,\Box = \Box_E \to \Box + i \vp$ prescription, the UV 
form factor  (\ref{kWinUV}) gains an imaginary addition. 
The imaginary term is known to describe the creation of 
massless particles by the gravitational field, as discussed in 
Refs.~\cite{ZS-77} (see also \cite{DobMar} for a more 
detailed derivation). This result shows the relation between 
conformal anomaly and the rate of particle creation, which is
(in the leading approximation)  proportional to the 
Weyl-squared and, to the $R^2$-squared terms in the 
nonlocal effective action. It would be interesting to extend
this result for the creation of massive particles in the very 
early Universe using the pseudoeuclidean analog of the 
form factors (\ref{45W})  and (\ref{45R}), which can be 
also easily generalized for the massive fermions and vectors
\cite{fervi}. We leave this investigation for the possible 
future work. 

Similar derivation for the Gauss-Bonnet part of the anomaly
is impossible, exactly because of the corresponding form factor
is of the third order in curvature, and therefore is beyond the
scope of the present consideration. The calculation of this
term in the strictly massless theory has been done in
\cite{bavi90-3order} and \cite{bavi-anom}.  At the same time,
the form factors are very helpful in better understanding the
problem of ambiguity of the conformal anomaly, related to
the local $R^2$-term in the anomaly-induced action and
$\cx R$-term in the UV divergence \cite{anomaly-2004}.
In particular, using  covariant Pauli-Villars regularization one
can show that this ambiguity takes place not only in the
dimensional \cite{duff94,birdav}, but also in other regularizations.
Moreover, if the conformal limit is achieved by taking the massless
limit $\xi \to 1/6$, $m\to 0$ in the non-conformal model, the
$R^2$-term remains non-local until the limit is taken, and then
there is no discontinuity or ambiguity in the anomaly-induced
result for the loop contribution.

In the IR limit, when $k^2 \ll m^2$, one can observe a very
different situation. The asymptotic behavior of $Y$ and $k_W$
is, in this case, of the power-like form, e.g.
\beq
Y &=& -\frac{1}{12}\,\frac{k^2}{m^2}\Big( 1\,-\,
\frac{1}{10}\,\frac{k^2}{m^2}\Big)\,+\,...
\label{AinIR}
\\
k_W &=& -\frac{1}{840}\,\frac{k^2}{m^2}\Big(1\,+\,
\frac{1}{18}\,\frac{k^2}{m^2}\Big)\,+\,...
\label{KWinIR}
\eeq
One can see that there is no logarithmic ``running'' in the IR and
hence there is no direct relationship between the dependence on
momenta and on $\mu$ in this region. This is the gravitational
decoupling, which can be also seen in the $\be$-functions \cite{apco}.
In other words, in the IR, nonlocal terms simply disappear, while
the divergences remain the same. Thus, the logarithmic divergences
give important glues on the UV behavior of the theory but
become non-informative in the IR.

\section{Conclusions}
\label{Conc}

We have presented a detailed derivation of the nonlocal form factor
in the scalar massive theory in curved spacetime. The logarithmically
divergent part does not depend on the regularization, the same
concerns the finite non-local part.

Let us briefly discuss some of the main unsolved problems
related to the form factors.

In the recent work \cite{324} the well-known result for the
form factors from the mixed diagrams (see e.g. \cite{Ilisie}) has
been generalized for the weak gravitational background. This type
of calculations is interesting, as it enables one to explore the IR
decoupling of the massive degrees of freedom in the renormalizable
models of quantum gravity, which have higher derivatives and
hence massive ghost-like degrees of freedom. Such a calculation
would confirm that Einstein's GR is the universal effective
model of quantum gravity in the IR \cite{don}.  Indeed, this
calculation meets serious technical difficulties and it would be
more than useful to work with the functional methods instead of
the diagrams.

Another interesting, albeit difficult problem, is to extend the
integration of the heat kernel \cite{apco,fervi} in the massive
theories, to the third-order in curvatures form factors, using the
general result of \cite{bavi90-3order}. Since in this case there
are several external momenta, the study of decoupling maybe
more interesting, especially because the third-order terms may
be relevant for the creation of particles from vacuum.

In conclusion, both diagrammatic and functional derivations
of nonlocal parts of effective action are, in general,
well-developed. However, there are new applications in this
area, which may be a challenging problems for the future.

\section*{Acknowledgements}

This work of I.Sh. was partially supported by Conselho Nacional de
Desenvolvimento Cient\'{i}fico e Tecnol\'{o}gico - CNPq under the
grant 303635/2018-5 and Funda\c{c}\~{a}o de Amparo \`a Pesquisa
de Minas Gerais - FAPEMIG under the project APQ-01205-16.



\begin {thebibliography}{99}

\bibitem{UtDW} R.~Utiyama and B.S.~DeWitt,
{\it Renormalization of a classical gravitational field
interacting with quantized matter fields,}
J. Math. Phys. {\bf 3} (1962) 608.

\bibitem{ZeldStar71}	Ya.B. Zeldovich and A.A. Starobinsky,
{\it Particle production and vacuum polarization in an anisotropic
gravitational field,}
Sov. Phys. JETP {\bf 34} (1972) 1159 
[Zh. Eksp. Teor. Fiz. {\bf 61} (1971) 2161]. 

\bibitem{apco} E.V. Gorbar and I.L. Shapiro,
{\it Renormalization group and decoupling in curved space,}
JHEP {\bf 02} (2003) 021, hep-ph/0210388.

\bibitem{omar2013} A. Codello and O. Zanusso,
{\it On the non-local heat kernel expansion,}
J. Math. Phys. {\bf 54} (2013) 013513, arXiv:1203.2034.

\bibitem{BunPar} T.S. Bunch and L. Parker,
{\it Feynman Propagator in Curved Space-Time: A Momentum
Space Representation.}
Phys. Rev. {\bf D20} (1979) 2499.

\bibitem{CorPot}  F. Sobreira, B.J. Ribeiro, and I.L. Shapiro,
{\it Effective Potential in Curved Space and Cut-Off Regularizations}.
Phys. Lett. {\bf B705} (2011) 273, 
arXiv: 1107.2262. 

\bibitem{BAES} I.L. Buchbinder, A. Rairis Rodrigues,
E.A. dos Reis and I.L. Shapiro,
{\it Quantum aspects of Yukawa model with scalar and axial
scalar fields in curved spacetime,}
Eur. Phys. J. {\bf C79} (2019) 1002,
arXiv:1910.01731.

\bibitem{DeWitt} B.S. DeWitt,
{\it Dynamical Theory of Groups and Fields,}
(Gordon and Breach, 1965).

\bibitem{bavi85}  A.O. Barvinsky and G.A. Vilkovisky,
{\it The generalized Schwinger-DeWitt technique in gauge theories and
quantum gravity,} Phys. Rep. {\bf 119}, (1985) 1.

\bibitem{avram} I.G. Avramidi,
{\it Covariant methods for the calculation of the effective action
in quantum field theory and investigation of higher-derivative
quantum gravity,} (PhD thesis, Moscow University, 1986).
hep-th/9510140;
{\it Covariant studies of nonlocal structure of effective action,}
Sov. J. Nucl. Phys.  {\bf 49}, 735 (1989)
[Yad. Fiz.  {\bf 49}, 1185 (1989), in Russian];
{\it Heat kernel and quantum gravity}, (Springer-Verlag, 2000).

\bibitem{bavi90} A.O.~Barvinsky and G.A.~Vilkovisky,
{\it Covariant perturbation theory. 2: Second order in the curvature.
General algorithms,}
Nucl. Phys. {\bf 333B} (1990) 471.

\bibitem{fervi} E.V. Gorbar and I.L. Shapiro,
{\it Renormalization group and decoupling in curved space:
II. The Standard Model and Beyond,}
JHEP {\bf 06} (2003) 004, hep-ph/0303124.

\bibitem{sebastian}
S.A.~Franchino-Viñas, T.~de Paula Netto, I.L.~Shapiro
and O.~Zanusso,
{\it Form factors and decoupling of matter fields in four-dimensional
gravity,}
Phys. Lett. {\bf B790} (2019) 229,
arXiv:1812.00460.

\bibitem{Bexi}
G. de Berredo-Peixoto, E.V. Gorbar, I.L. Shapiro,
{\it On the renormalization group for the interacting massive
scalar field theory in curved space,}
Class. Quant. Grav. {\bf 21} (2004) 2281, 
hep-th/0311229.

\bibitem{Shocom} I.L. Shapiro, J. Sol\`{a},
{\it Massive fields temper anomaly-induced inflation,}
Phys. Lett. {\bf B530} (2002) 10, 
hep-ph/0104182.

\bibitem{asta} A.M. Pelinson, I.L. Shapiro and F.I. Takakura,
{\it  On the stability of the anomaly-induced inflation,}
Nucl. Phys. {\bf B648} (2003) 417, 
hep-ph/0208184.

\bibitem{AC} T.~Appelquist and J.~Carazzone,
{\it Infrared Singularities and Massive Fields,}
Phys. Rev.  {\bf D11} (1975) 2856.

\bibitem{GuZeln}  Y.V.~Gusev and A.I.~Zelnikov,
{\it Finite temperature nonlocal effective action for quantum
fields in curved space,}
Phys. Rev. {\bf D59} (1999) 024002,
hep-th/9807038.

\bibitem{Salam51} A.~Salam,
{\it Divergent Integrals in Renormalizable Field Theories,}
Phys. Rev. {\bf 84} (1951) 426.

\bibitem{Liao} S.-B. Liao,
{\it Connection between momentum cutoff and operator cutoff
regularizations,}
Phys. Rev. {\bf D53} (1996) 2020.

\bibitem{leibr} G. Leibbrandt,
{\it Introduction to the technique of dimensional regularization,}
Mod. Phys. Rep. {\bf 47} (1975) 849. 

\bibitem{nelspan82} B.L. Nelson and P. Panangaden,
{\it Scaling Behavior Of Interacting Quantum Fields In
Curved Space-Time,}
Phys.Rev. {\bf D25} (1982) 1019.

\bibitem{tmf} I.L. Buchbinder,
{\it On Renormalization group equations in curved space-time,}
Theor. Math. Phys. {\bf 61} (1984) 393.

\bibitem{book}
I. L. Buchbinder, S. D. Odintsov e I. L. Shapiro,
{\it Effective action in Quantum Gravity}, (IOP Publishing,
Bristol, 1992).

\bibitem{highderi} M. Asorey, J.L. L\'opez and I.L. Shapiro,
{\it Some remarks on high derivative quantum gravity,}
Int. Journ. Mod. Phys. {\bf A12} (1997) 5711.

\bibitem{PoImpo} I.L.~Shapiro,
{\it Effective action of vacuum: semiclassical approach},
Class. Quant. Grav. {\bf 25} (2008) 103001,
arXiv:0801.0216.

\bibitem{ZS-77} Ya. B. Zeldovich and A. A. Starobinsky,
{\it Rate of particle production in gravitational fields,}
JETP Lett. {\bf 26} (1977) 252.

\bibitem{DobMar} A.~Dobado and A.L.~Maroto,
{\it Particle production from nonlocal gravitational
effective action,}
Phys. Rev. \textbf{D60} (1999) 104045,
gr-qc/9803076.

\bibitem{bavi90-3order} A.O.~Barvinsky and G.A.~Vilkovisky,
{\it Covariant perturbation theory. 3: Spectral representations
of the third order form-factors,}
Nucl. Phys. {\bf B333} (1990) 512.

\bibitem{bavi-anom} A.O. Barvinsky, Yu.V. Gusev,
G.A. Vilkovisky and V.V. Zhitnikov,
{\it The One loop effective action and trace anomaly in
four-dimensions,} Nucl.Phys. {\bf B439} (1995) 561,
hep-th/9404187.

\bibitem{anomaly-2004} M. Asorey, E.V. Gorbar and I.L. Shapiro,
{\it Universality and Ambiguities of the Conformal Anomaly,}
Class. Quant. Grav. {\bf 21} (2004) 163.

\bibitem{duff94} M.J. Duff,
{\it Twenty years of the Weyl anomaly,}
Class. Quant. Grav. {\bf 11} (1994) 1387,
hep-th/9308075.

\bibitem{birdav} N.D. Birell and P.C.W. Davies, {\it Quantum fields
in curved space}, (Cambridge Univ. Press, Cambridge, 1982).

\bibitem{324} T.G. Ribeiro and I.L. Shapiro,
{\it Scalar model of effective field theory in curved space,}
JHEP {\bf 1910} (2019) 163,
\ arXiv:1908.01937.

\bibitem{Ilisie} V. Ilisie, {\it Concepts in Quantum Field Theory.
A Practitioner's Toolkit,} (Springer, 2016).

\bibitem{don} J.F. Donoghue,
{\it Leading quantum correction to the Newtonian potential,}
Phys. Rev. Lett. {\it 72} (1994) 2996, 
gr-qc/9310024; \ \
{ \it General relativity as an effective field theory:
The leading quantum corrections},
Phys. Rev. {\bf D50} (1994) 3874, 
gr-qc/9405057.



\end{thebibliography}
\end{document}